\documentclass[12pt]{article}
\usepackage{amssymb}
\usepackage{epsfig}
\usepackage{sint,cite}

\newcommand\be{\begin{equation}}
\newcommand\ee{\end{equation}}
\newcommand\bea{\begin{eqnarray}}
\newcommand\eea{\end{eqnarray}}
\newcommand\ba{\begin{array}}
\newcommand\ea{\end{array}}
\newcommand{\rmO}{{\rm O}}
\def\rmA{{\rm A}}
\def\rmR{{\rm R}}
\def\rmP{{\rm P}}
\def\ZP{Z_{\rmP}}
\def\sp{\sigma_{\rm P}}
\def\SP{\Sigma_{\rm P}}
\def\ZA{Z_{\rmA}}
\def\ZAc{Z_{\rmA}^{\,\rm con}}
\def\zM{Z_{\rm M}}
\def\nf{N_{\rm f}}
\newcommand{\fP}{f_{\rm P}}
\newcommand{\fA}{f_{\rm A}}
\newcommand{\cA}{c_{\rm A}}
\newcommand{\csw}{c_{\rm sw}}
\def\ct{c_{\rm t}}

\def\gbar{\bar{g}}
\def\mbar{\overline{m}}
\def\mbars{\mbar_{\rm s}}
\def\Mstrange{M_{\rm s}}
\def\Mref{M_{\rm ref}}
\def\mref{m_{\rm ref}}

\newcommand{\eq}[1]{Eq.~(\ref{#1})}
\newcommand{\fig}[1]{Fig.~\ref{#1}}
\newcommand{\tab}[1]{Table~\ref{#1}}
\newcommand{\sect}[1]{Section~\ref{#1}}

\def\fm{\,{\rm fm}}
\def\Lmax{L_{\rm max}}

\def\MSb{\overline{{\rm MS}}}
\def\GeV{\,{\rm GeV}}
\def\MeV{\,{\rm MeV}}
\def\fm{\,{\rm fm}}
\def\Nf{N_{\rm f}}
\def\mps{m_{\rm PS}}
\def\kref{\kappa_{\rm ref}}
\def\mref{m_{\rm ref}}
\def\Mref{M_{\rm ref}}

\def\Mu{M_{\rm u}}

\def\Md{M_{\rm d}}
\def\ms{m_{\rm s}}
\def\Ms{M_{\rm s}}
\def\dd{{\rm d}}
\newcommand{\Ffrac}[2]{\frac{\displaystyle{#1}}{\displaystyle{#2}}}
\def\psibar{\overline{\psi}}
\def\bA{b_{\rm A}}
\def\bP{b_{\rm P}}
\def\mq{m_{\rm q}}
\def\mK{m_{\rm K}}
\def\FK{F_{\rm K}}
\def\DH{\Delta H}
\def\DHc{|\DH({\rm cycle})|}
\def\BRGI{B_{\rm RGI}}

\def\ma[#1,#2,#3,#4]  {{\left( \matrix{ #1  & #2 \cr
                                        #3  & #4 \cr } \right)}}

\begin{document}

\thispagestyle{empty}
\title{{\normalsize\vskip -50pt
\mbox{} \hfill DESY 05-124 \\
\mbox{} \hfill HU-EP-05/31 \\
\mbox{} \hfill SFB/CPP-05-32 \\}
\vskip 25pt
Non--perturbative quark mass renormalization in two--flavor QCD
}

\author{
\centerline{
            \epsfxsize=2.5 true cm
            \epsfbox{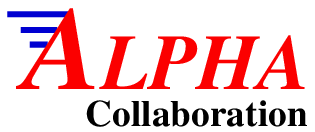}}\\
Michele Della Morte $^{a}$, Roland Hoffmann $^{a}$, Francesco Knechtli $^{a}$, \\
Juri Rolf $^{a}$, Rainer Sommer $^{b}$, Ines Wetzorke $^{c}$ and Ulli Wolff $^{a}$
\\[0.5cm]
$^{a}$ Institut f\"ur Physik, Humboldt Universit\"at,\\
Newtonstr. 15, 12489 Berlin, Germany\\[0.25cm]
$^{b}$ DESY, Platanenallee 6, 15738 Zeuthen, Germany\\[0.25cm]
$^{c}$ John von Neumann--Institut f\"ur Computing NIC,\\ 
Platanenallee 6, 15738 Zeuthen, Germany\\[0.5cm]
}
\date{}

\maketitle

\begin{abstract}
The running of renormalized quark masses is computed in lattice QCD 
with two flavors of massless O($a$) improved Wilson quarks.
The regularization and flavor independent factor that relates running
quark masses to the renormalization group invariant ones is evaluated
in the Schr{\"o}dinger Functional scheme.
Using existing data for the scale $r_0$ and the pseudoscalar meson masses,
we define a reference quark mass
in QCD with two degenerate quark flavors.
We then compute the renormalization group invariant reference quark mass
at three different lattice spacings.
Our estimate for the continuum value is converted to 
the strange quark mass with the help of chiral perturbation theory.
\end{abstract}

%\begin{flushright}
%  DESY 05-xxx \\
%  HU-EP-05/xx \\
%  SFB/CPP-05-xx
%\end{flushright}

\newpage

%%% Local Variables: 
%%% mode: latex
%%% TeX-master: "mbar"
%%% End: 

\section{Introduction}
\label{introduction}

Lattice QCD provides a definition of quark masses from first principles.
For example
given a hadronic input accessible to experiments, such as the K--meson mass $\mK$
and decay constant $\FK$, it is possible to compute the strange quark mass
on the lattice. A convenient quantity to consider is the renormalization group
invariant (RGI) strange quark mass,
which is independent of the renormalization scheme
if the renormalization conditions are imposed at zero quark mass
\cite{Weinberg:1951ss}.

If one looks at the most recent lattice results for the strange quark mass,
from simulations with two
\cite{AliKhan:2001tx,Aoki:2002uc,Gockeler:2004rp} and two plus one
\cite{Ishikawa:2004xq,Aubin:2004ck}
dynamical quark-flavours, there is a large spread of values ranging from 
68$\MeV$ to 132$\MeV$. There are several sources of
systematic errors in these computations, such as the use of perturbative
renormalization (except for \cite{Gockeler:2004rp}, where non--perturbative
renormalization is done in the RI--MOM scheme) and the values of the
lattice spacing affordable nowadays.

To perform a completely controlled computation we start from the 
bare PCAC mass $m_i(g_0)$ for a given quark flavor $i$ and
determine the RGI mass
\be
   M_i = \zM(g_0) \, m_i(g_0) \,. \label{e:zM}
\ee
The bare coupling $g_0$ is in a one--to--one relation to
the lattice spacing and the continuum
limit of $M_i$ exists (and should be taken). 
As an intermediate step we first define the running mass
\be
\mbar_i(\mu)={\ZA(g_0) \over \ZP(g_0,a\mu)}\, m_i(g_0) \,,
\label{mbarintro}
\ee
where $\mbar$ is non--perturbatively defined in
the Schr{\"o}dinger Functional renormalization scheme and hence
is well--defined also for low energies $\mu$.
This corresponds to splitting $\zM$ into two factors
\be
   \zM(g_0) = {M \over \mbar(\mu)} \times {\ZA(g_0) \over \ZP(g_0,a\mu)} \,,
   \label{e:zMsplit}
\ee
which will be computed with full non--perturbative precision following the
strategy of \cite{Capitani:1998mq}. Note that all renormalization factors
are flavor--independent and we hence omit the subscript $i$ in mass ratios.
In the first factor of the above splitting the universal continuum limit
is understood to have been taken. Its computation is the main objective of
this paper. This result can then be used for 
any action and only the second factor needs to be
redetermined. This becomes tractable 
by making the universal factor
available for $\mu$ in the range of hadronic energies of $\rmO(1-2 \GeV)$.
As for the running coupling \cite{DellaMorte:2004bc} our method thus 
avoids the need to treat a multi--scale problem in a large volume
in this step.

First results on the $\mu$--dependence of $\mbar$ in the $\nf=2$ theory
have already appeared in Refs.
\cite{Knechtli:2002vp,Knechtli:2003mq,DellaMorte:2003jj}.
Beyond a finalization of these results, we here compute the
second factor --- and hence $\zM$ --- using non--perturbative improvement
\cite{DellaMorte:2005se} and renormalization \cite{DellaMorte:2005rd} of the
axial current in the theory with two flavors of O($a$) improved Wilson
quarks and plaquette gauge action. 

As an application, one wants to determine the light (up, down, strange)
quark masses.\footnote{Later, when lattice spacings are small
enough, the charm quark mass
can be determined \cite{Rolf:2002gu,Dougall:2004hx}. Also the beauty quark mass
computed in HQET through the strategy of \cite{Heitger:2003nj}, 
is based on \eq{e:zM} in QCD.}
For practical reasons this is at present not yet possible for us
in the most straight-forward way by simulations at the physical parameters.
We need a hierarchy of additional approximations to compute for
instance the strange quark mass.

First of all our simulation algorithm is at the moment still restricted
to pairs of degenerate flavours and we include one such pair ($\Nf=2$).
In the large volume simulations needed to determine $m_i(g_0)$ in \eq{e:zM}
our masses can at present not  be taken to values 
small enough for the up-- and down--flavors, for instance
by tuning the pseudoscalar states to the physical pion mass. Instead
we shall determine the RGI quark mass $\Mref$ that is associated
with a ``Kaon'' made from two degenerate flavours.

The analogous computation of $\Mref$ has been performed
in the quenched \mbox{$\Nf=0$} theory \cite{Garden:1999fg}
with a result that agrees within errors
with the new one. We therefore assume that at the present level of accuracy
it also applies to a hypothetical QCD with three degenerate flavors.
A relation between this model and real QCD is finally established 
by chiral perturbation theory \cite{Gasser:1984gg,Leutwyler:1996qg} supplemented 
with some knowledge \cite{Leutwyler:1996qg,Amoros:2001cp,Wittig:2002ux}
of the phenomenologically inaccessible \cite{Kaplan:1986ru}
low energy constants. The conclusion of \cite{Leutwyler:1996qg}
is that the ratio of quark masses is close to the 
one given at lowest order in chiral perturbation theory. 
Given this ``fact'' --- but keeping in mind that it
should be scrutinized in future lattice QCD computations ---  
it is then sufficient to compute any one quark mass
from lattice QCD. In particular we can connect our $\Mref$
with the strange mass $\Mstrange$.
The result that corresponds to the Gell-Mann--Oakes--Renner formula
\cite{Gell-Mann:1968rz} is $\Mstrange = 48/25 \Mref$.

Finally, the conversion of the RGI mass to the conventionally cited 
$\MSb$ mass at 2~$\GeV$ renormalization scale is of course 
based on perturbation theory, which does however look
very well behaved, see \tab{t_moM}.

%%% Local Variables: 
%%% mode: latex
%%% TeX-master: "mbar"
%%% End: 

\section{The renormalization scheme}
\label{renormalization}

QCD is a theory which has as free parameters the bare gauge coupling $g_0$
and the bare quark masses $m_i\,,\;i=1,...,\Nf$.
The hadronic scales like $\FK$ or $\mK$ are connected to the
perturbative high energy regime of QCD via the running of
renormalized couplings
\bea
 \gbar^2(\mu) \,=\, Z_gg_0^2\,, & & 
 \mbar_i(\mu) \,=\, Z_mm_i\,,
\eea
where $\mu$ is the renormalization scale. 
In the following we assume that the renormalization conditions are 
imposed at {\em zero quark mass} (mass--independent schemes)
\cite{Weinberg:1951ss}.
Introducing a regularization prescription, e.g. a lattice spacing $a$, the
renormalization factors are functions of $g_0$ and $a\mu$
\bea
 Z_g \,=\, Z_g(g_0,a\mu) \,, & &
 Z_m \,=\, Z_m(g_0,a\mu) \,.
\eea
In renormalized quantities the regulator can be removed, e.g.
the continuum limit $a\to0$ can be taken, yielding a finite result.
The advantage of mass--independent renormalization schemes
is that in all such schemes
the {\em ratios} of renormalized quark masses 
for different flavors are scale and scheme independent constants
\cite{Leutwyler:1996qg,Sint:1998iq}.

The running of the renormalized couplings is described by the
renormalization group equations (RGE)
\bea
\mu\frac{\dd\gbar}{\dd\mu}=\beta(\gbar)\,, &&
\mu\frac{\dd\mbar_i}{\dd\mu}=\tau(\gbar)\mbar_i \,.
\eea
The $\beta$ and $\tau$ functions are non-perturbatively defined if this is true for
$\gbar$ and $\mbar_i$. Their perturbative expansions are
\bea
 \beta(\gbar) &
 _{\mbox{$\stackrel{\displaystyle\sim}{\scriptstyle \gbar\rightarrow0}$}} &
 -\gbar^3\{b_0+b_1\gbar^2+b_2\gbar^4+...\} \,, \label{betafunc} \\
 \tau(\gbar) &
 _{\mbox{$\stackrel{\displaystyle\sim}{\scriptstyle \gbar\rightarrow0}$}} &
 -\gbar^2\{d_0+d_1\gbar^2+...\} \,. \label{taufunc}
\eea
The coefficients
\bea
 b_0 \,=\, \frac{1}{(4\pi)^2}\left(11 - \frac{2}{3}\Nf\right) \,,\quad
 b_1 \,=\, \frac{1}{(4\pi)^4}\left(102 - \frac{38}{3}\Nf\right) \,,\quad
 d_0 \,=\, \frac{8}{(4\pi)^2}\,,
\eea
are scheme independent.
A physical quantity $P$ is a quantity for which
the total dependence on the renormalization scale 
$\mu$ vanishes, i.e. it is a renormalization group invariant (RGI)
\bea
 \mu\frac{\dd}{\dd\mu}P(\mu,\gbar,\{\mbar_i\}) & = & 0 \,.
\eea
Examples are the $\Lambda$--parameter and the RGI quark masses
\bea
 \Lambda & = & \mu(b_0\gbar^2)^{-b_1/2b_0^2}
 {\rm e}^{-1/(2b_0\gbar^2)}
 \exp\left\{-\int_0^{\gbar}\dd x\left[
 \Ffrac{1}{\beta(x)}+\Ffrac{1}{b_0x^3}-\Ffrac{b_1}{b_0^2x}\right]\right\}
 \,,\label{LAMBDA} \\
 M_i & = & \mbar_i(2b_0\gbar^2)^{-d_0/2b_0}
 \exp\left\{-\int_0^{\gbar}\dd x\left[
 \Ffrac{\tau(x)}{\beta(x)}-\Ffrac{d_0}{b_0x}\right]\right\}
 \,,\label{MRGI}
\eea
where $\gbar=\gbar(\mu)$ and $\mbar_i=\mbar_i(\mu)$. The
$\Lambda$--parameter and the RGI quark masses
are defined independent of perturbation theory and
their connections between different
mass independent renormalization schemes can be given in a simple and exact way
\cite{Sint:1998iq}. In particular
the RGI quark masses $M_i$ are {\em scheme independent}.
Any physical quantity $P$ can be considered to be a function of $\Lambda$ and $M_i$,
i.e. there exist a function $\hat{P}$ such that \cite{Sint:1998iq}
\bea
 P(\mu,\gbar,\{\mbar_i\}) & = & \hat{P}(\Lambda,\{M_i\}) \,.
\eea
For this reason $\Lambda$ and $M_i$ seem preferable as the fundamental parameters
of QCD.

\subsection{Quark masses in the Schr{\"o}dinger Functional \label{quarkSF}}

In QCD a renormalized mass is defined through the
partially conserved axial current (PCAC) 
relation, which involves the renormalized axial current $(A_{\rmR})_\mu(x)$
and the renormalized pseudoscalar density $P_{\rmR}(x)$
\bea
 \partial_{\mu}(A_{\rmR})_{\mu} & = & (\mbar_i+\mbar_j)P_{\rmR} \,,\\
 (A_{\rmR})_{\mu}(x) & = & \ZA\psibar_i(x)\gamma_{\mu}\gamma_5\psi_j(x) \,,\\
 P_{\rmR}(x) & = & \ZP\psibar_i(x)\gamma_5\psi_j(x) \,.
\eea
The renormalization constant $\ZA$ can be calculated non--perturbatively,
using chiral Ward identities and does not
depend on the renormalization scale.

The renormalization constant $\ZP$ can be conveniently determined in the 
Schr\"odinger Functional (SF) renormalization scheme
\cite{Jansen:1995ck,Luscher:1998pe}.
There QCD is formulated in a finite box of spatial size $L$ and temporal extent
$T$. The fields are subject to Dirichlet boundary conditions in time, which
provide an infrared cutoff to the frequency spectrum of quarks and gluons.
This allows to perform simulations at zero quark mass and thus to
use the SF as a mass--independent renormalization scheme.
Our renormalization scheme is
further specified by setting $T=L$ (see below). The renormalization conditions
are then naturally imposed at the scale $\mu=1/L$.

The presence of boundary values for the fields in the SF formulation of
a field theory requires in general additional
(compared to the case without boundaries)
renormalization \cite{Symanzik:1981wd,Luscher:1985iu}.
The renormalizability of QCD with SF boundaries has been studied in Refs.
\cite{Luscher:1992an,Sint:1993un,Sint:1995rb,Sint:1995ch} where it was shown
that no additional counterterms are needed except for one boundary term which
amounts to a rescaling of the boundary values of the fermion fields
by a logarithmically divergent factor.

\begin{figure}[t]
 \begin{center}
     \includegraphics[width=5cm]{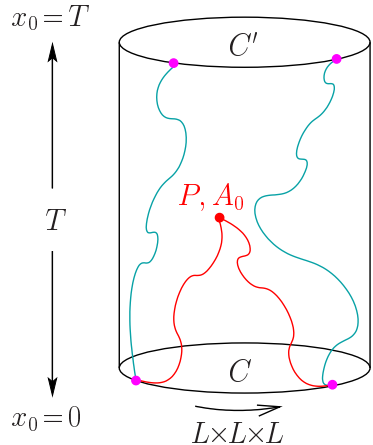}
 \end{center}
 \vspace{-0.5cm}
 \caption{The Schr\"odinger Functional.}
 \label{f_sf}
\end{figure}

The renormalization condition for $\ZP$ that we employ is discussed in Refs.
\cite{Jansen:1995ck,Sint:1998iq,Capitani:1998mq}.
It uses a correlation function $\fP(x_0)$,
which is a matrix element of the pseudoscalar density inserted at time distance
$x_0$ from a pseudoscalar boundary state, the other boundary state having
vacuum quantum numbers.
To cancel the multiplicative renormalization of the boundary quark fields
the boundary--to--boundary correlation function $f_1$ is used.
The renormalization constant $\ZP$ is then defined through
\bea
 \ZP = \frac{\sqrt{3f_1}}{f_{\rmP}(L/2)} \quad \mbox{at} \quad
 m_i=0\,,\;i=1,\ldots,\Nf\,. \label{ZPdef}
\eea
The correlation functions are schematically represented in \fig{f_sf}.
They are computed at zero quark masses $m_i$.
The definition \eq{ZPdef} is such that $\ZP=1$ at tree level of perturbation
theory.
The renormalization condition for $\ZP$ is further specified by setting
\bea
 T\,=\,L\,,\quad & C=C^{\prime}=0\,,\quad & \theta\,=\,0.5\,, \label{SFparams}
\eea
where $C\,,\,C^{\prime}$ are the boundary gauge fields in the Lie algebra
at $x_0=0$ and $x_0=T$ and
$\theta$ the parameter controlling the spatial boundary conditions of the
fermion fields. For more details of the calculation we refer to 
\cite{Capitani:1998mq,Sint:1998iq}.

A rigorous definition of the renormalized mass can be given in the lattice
regularization of QCD. In our case we work with $\Nf=2$ mass--degenerate flavors of
O($a$) improved \cite{Luscher:1996sc,Jansen:1998mx} Wilson fermions.
The massless theory is defined in the bare parameter space along the line
$\kappa=\kappa_c(g_0)$ where the PCAC mass
\bea
 m(g_0,\kappa) & = & 
               \left. \frac{\frac{1}{2}(\partial_0^{\ast}+\partial_0)\fA(x_0)
                +\cA a\partial_0^{\ast}\partial_0\fP(x_0)}
               {2\fP(x_0)}\right|_{x_0=T/2} \label{PCACmass}
\eea
vanishes.
Here, $\partial_0$ and $\partial_0^{\ast}$ are the forward and backward lattice
derivatives, respectively.
The O($a$) improvement coefficient $\cA$ of the axial current
has been computed non--perturbatively in Ref. \cite{DellaMorte:2005se}.
The correlation function $\fA(x_0)$ of the axial current $A_0$ is defined
analogously to $\fP(x_0)$.
The renormalized PCAC mass at the scale $\mu=1/L$ can then be written as
\bea
 \mbar(\mu) \,=\, \left.\lim_{a\to0}\,
 Z_m(g_0,a\mu)\,m(g_0,\kappa)\right|_{u=\gbar^2(L)} \,,\quad
 Z_m(g_0,a\mu)\,=\,\frac{\ZA(g_0)}{\ZP(g_0,L/a)} \,,\label{mbar}
\eea
where the renormalized gauge coupling $\gbar^2(L)$ \cite{DellaMorte:2004bc}
is kept fixed. Since the coupling runs with the box size $L$, keeping $\gbar^2$
fixed means keeping the renormalization scale $\mu$ fixed.
The renormalization factor $\ZA$ of the axial current has been
determined non--perturbatively in Ref. \cite{DellaMorte:2005rd}.
The precise definition of $\ZP$ (differing at order $a^2$ from \eq{ZPdef}) is
\bea
 \ZP(g_0,L/a) & = & c\frac{\sqrt{3f_1}}{f_{\rmP}(L/2)} \quad 
 \mbox{at} \quad \kappa=\kappa_c\,, \label{ZPlat}
\eea
where the factor $c(L/a)$ is chosen such that $Z_P(0,L/a)=1$ and can be found in
\cite{Capitani:1998mq}.

In order that the continuum limit
is reached with cutoff effects strictly proportional to $a^2$, 
the $O(a)$ improvement factor $(1+(\bA-\bP)a\mq)$ should be included in \eq{mbar}
\cite{Luscher:1996sc}.
Here $a\mq=(1/\kappa-1/\kappa_c)/2$ is the bare subtracted quark mass.
Results in perturbation theory \cite{Sint:1997jx} and in
the quenched approximation \cite{Guagnelli:2000jw} show that
the difference of improvement coefficients $\bA-\bP$ is small.
The quark masses in our simulations will also be relatively small and we expect
corrections due to $\bA-\bP$ at the per mille level, which we neglect.

\section{The running of the mass in the SF--scheme}
\label{running}

\begin{table}[t] 
 \centering
  \begin{tabular}{lll}
   \hline\\
   \multicolumn{1}{c}{$u$}  & 
   \multicolumn{1}{c}{$\sp(u)$} & 
   \multicolumn{1}{c}{$\chi^2/n_{\rm df}$} \\[1.0ex]
   \hline\\
0.9793 & 0.9654(9)(11) & 2.16 \\
1.1814 & 0.9527(11)(6) & 0.47 \\
1.5031 & 0.9413(16)(2) & 0.01 \\
2.0142 & 0.9174(16)(24) & 3.58 \\
2.4792 & 0.8871(23)(18) & 0.54 \\
3.3340 & 0.8384(35)(12) & 0.20 \\[1.0ex]
   \hline
  \end{tabular} 
 \caption{Continuum extrapolations of $\SP$ fitting the $L/a=8$ and $L/a=12$
 data to a constant. The first error is statistical. The second error is
 the difference between the fit and the $L/a=8$ results and
 will be added linearly as a systematic error.}
 \label{t_sigmaP}
\end{table}
\begin{figure}[t]
 \begin{center}
     \includegraphics[width=11cm]{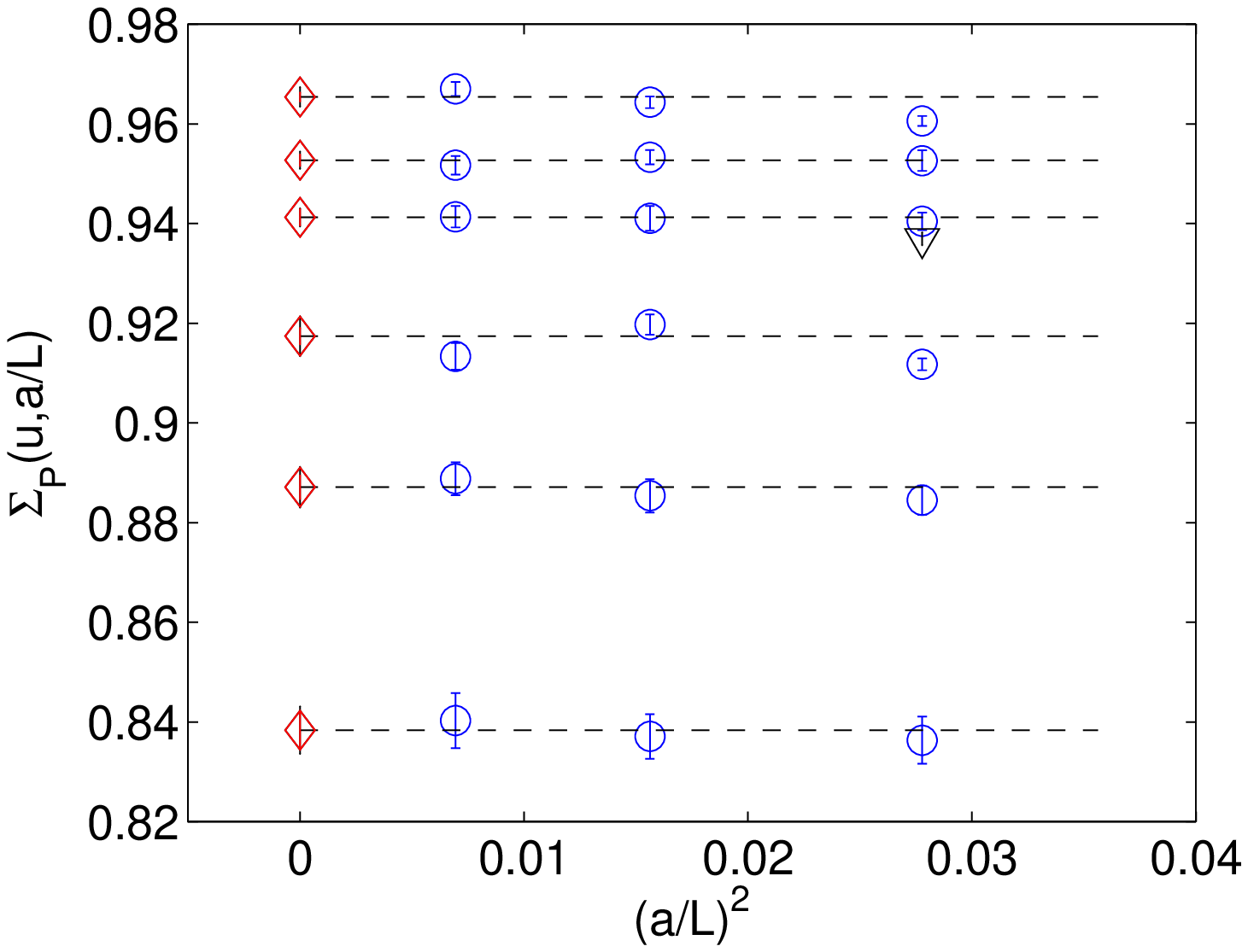}
 \end{center}
 \vspace{-0.5cm}
 \caption{Continuum extrapolations of $\SP$. The $L/a=6$ data have been excluded
 from the fits. For the third smallest coupling, the two points at $L/a=6$ refer
 to 1--loop and 2--loop values for the boundary improvement coefficient $c_t$
 \cite{Luscher:1992an,Bode:1999sm}.}
 \label{f_SigmaP}
\end{figure}

The running of the renormalized quark mass $\mbar(\mu)$
in the SF scheme as specified in \sect{quarkSF}
with $\Nf=2$ mass--degenerate flavors
can be computed on the lattice from
the step scaling function of the renormalization factor $\ZP$
extrapolated to the continuum
\bea
 \sp(u) & = & \left.\lim_{a\to0}\,\SP(u,a/L) \,,\quad
 \SP(u,a/L) \,=\,\frac{\ZP(g_0,2L/a)}{\ZP(g_0,L/a)}\right|_{u=\gbar^2(L)} 
 \label{ssfZP}\,.
\eea
From \eq{mbar} it follows immediately that
\bea
 \sp(u) & = & \frac{\mbar(\mu)}{\mbar(\mu/2)} \quad \mbox{for $\mu=1/L$} \,,
 \label{ssfmbar}
\eea
i.e. the step scaling function $\sp(u)$ describes the running of the renormalized
quark mass. We computed
$\SP(u,a/L)$ at six values of the renormalized coupling
$u$ corresponding approximately to a range of box sizes of the order
$L=10^{-2}\fm\ldots1\fm$ (or equivalently $\mu$ of the order $100\GeV\ldots1\GeV$).
At each value of $u$ we simulated
three lattice resolutions $L/a=6,8,12$ and the results for $\ZP$ and $\SP$ are
summarized in \tab{t_SigmaP} in Appendix \ref{app_zp}.
For the extrapolation to the continuum, we fitted to a constant
the two values of $\SP$ on the finer lattices, separately for each coupling $u$.
We then added linearly the difference between the fit and the $L/a=8$ result 
as a systematic error. The continuum estimates can be seen in \fig{f_SigmaP}.
Our data do not show any significant dependence on the lattice spacing,
as we could verify by trying different extrapolations (quadratic, linear in $a$).
This statement is based on the statistical accuracy that we could achieve.
We remark that also in the quenched approximation the cutoff effects were found
to be small \cite{Capitani:1998mq} and there $\Sigma_P$ was computed at an
even finer lattice resolution $L/a=16$.
The continuum values of $\sp(u)$ are summarized in \tab{t_sigmaP}. In the last 
column we list the $\chi^2$ divided by the number of degrees of freedom
$n_{\rm df}$ of the fit. Their average is close to the expected value of one.
\begin{figure}[t]
 \begin{center}
     \includegraphics[width=11cm]{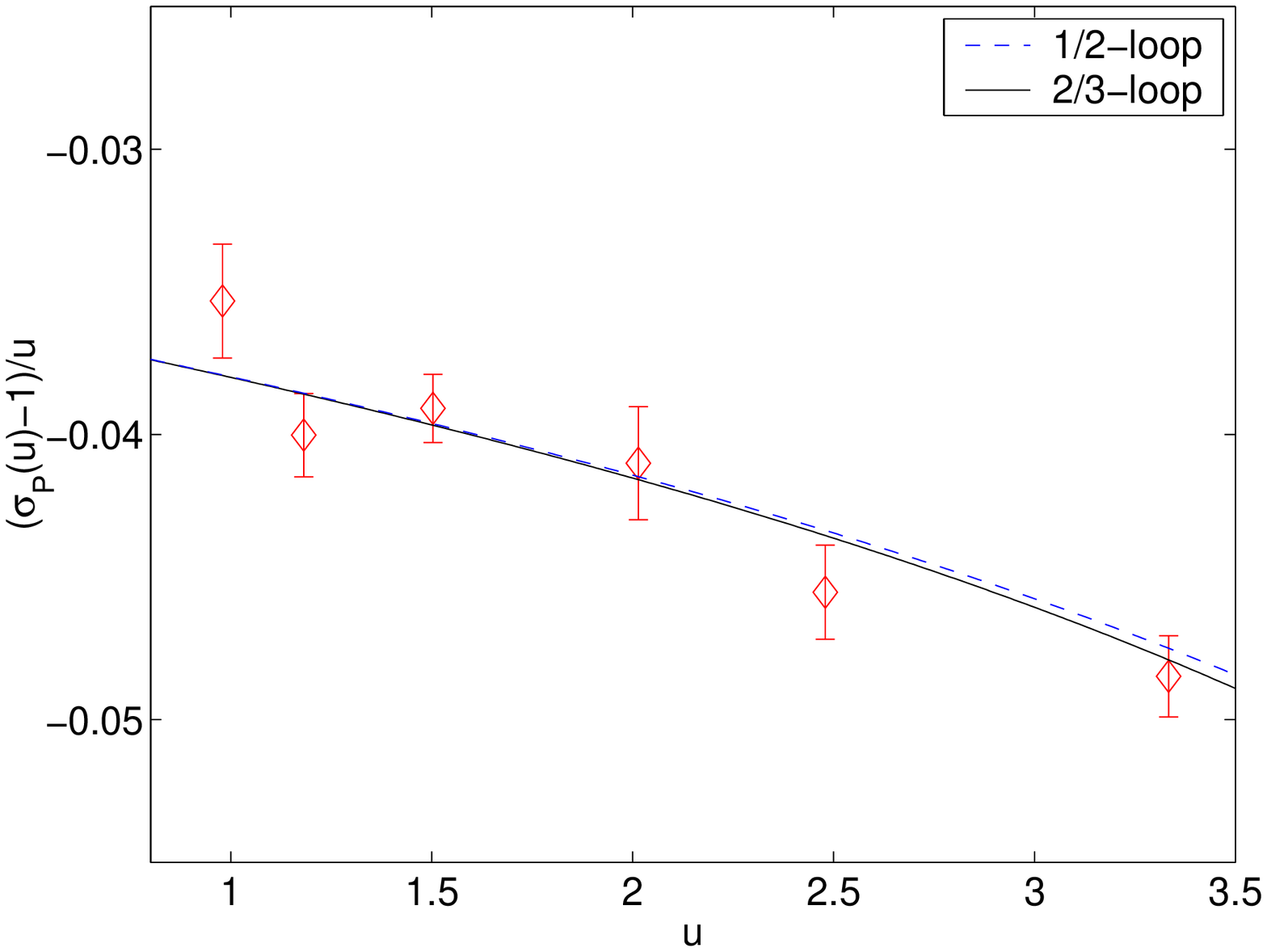}
 \end{center}
 \vspace{-0.5cm}
 \caption{Comparison of the non--perturbative data for the step scaling
 function $\sp(u)$ with perturbation theory. 2/3--loop refers to the 2--loop
 $\tau$-function and 3--loop $\beta$--function, analogously 1/2--loop.}
 \label{f_sigmaP}
\end{figure}

In perturbation theory the step scaling function $\sp(u)$ has an expansion
$\sp(u) = 1 - \ln(2)d_0u + {\rm O}(u^2)$. In \fig{f_sigmaP} our
non--perturbative data for $\sp(u)$ are conveniently plotted for comparison
with perturbation theory. We relate $\sp(u)$ to $\tau$ and $\beta$ using \eq{MRGI}
\bea
 \sp(u) & = & \left(\frac{u}{\sigma(u)}\right)^{d_0/(2b_0)}\exp\left\{
 -\int_{\sqrt{u}}^{\sqrt{\sigma(u)}}
 \dd x\left[\frac{\tau(x)}{\beta(x)}-\frac{d_0}{b_0x}\right]\right\} \,,
 \label{ptsp1}
\eea
where $\sigma(u)$ is the step scaling function of the coupling and
is determined by \cite{DellaMorte:2004bc}
\bea
 -2\ln(2) & = & \int_u^{\sigma(u)}{\rm d}x\,\frac{1}{\sqrt{x}\beta(\sqrt{x})}
 \,. \label{pts1}
\eea
Using for the $\tau$--function the 2--loop expression
with \cite{Sint:1998iq}
\bea
 d_1 & = & d_0 (0.0271 + 0.0105\Nf)
\eea
and for the $\beta$--function the 3--loop expression
with (see \cite{DellaMorte:2004bc})
\bea
 b_2 & = & (0.483 - 0.275\Nf + 0.0361\Nf^2 - 0.00175\Nf^3)/(4\pi)^3
\eea
we get from \eq{ptsp1} and \eq{pts1}
the perturbative curve shown in \fig{f_sigmaP}. Our non--perturbative data do
not show any significant deviation from the perturbative estimates.

We now take the continuum values for $\sp(u)$ in \tab{t_sigmaP} 
(with added statistical and systematic errors) and for
the step scaling function of the coupling $\sigma(u)$ the results 
from fits to constants in Table 4 of Ref. \cite{DellaMorte:2004bc}.
We solve the following joint recursion to evolve coupling and mass
from a low energy scale $1/\Lmax$ defined by
\bea
 u_0 \,=\,\gbar^2(\Lmax) & = & 4.61
\eea
to the higher scales $1/L_k$, $k=0,1,\ldots,8$ (with $L_0\equiv\Lmax$)
\bea
\left\{\ba{l} u_0=\gbar^2(\Lmax)=4.61 \\ \sigma(u_{k+1})=u_k \ea\right.
& \Rightarrow & u_k=\gbar^2(L_k) \,,\quad L_k = 2^{-k}\Lmax \,, \label{ucoeff} \\
\left\{\ba{l} w_0=1 \\ w_k=\left[\prod_{i=1}^k\sp(u_i)\right]^{-1} \ea\right.
& \Rightarrow & w_k=\frac{\mbar(1/\Lmax)}{\mbar(1/L_k)} \,. \label{wcoeff}
\eea
We interpolate the values of $\sigma(u)$ and $\sp(u)$
through a polynomial ansatz
\bea
 \sigma(u) & = & u + s_0u^2 + s_1u^3 + s_2u^4 + s_3u^5 + s_4u^6 \,, \label{pts2} \\
 \sp(u) & = & 1 + p_0u + p_1u^2 + p_2u^3 \,, \label{ptsp2}
\eea
where the coefficients $s_0$, $s_1$ \cite{DellaMorte:2004bc}
and $p_0=- \ln(2)d_0$ are fixed to their perturbative values.
The coefficients $s_2,\,s_3,\,s_4$ and $p_1,\,p_2$ are here fit parameters. 
The errors of the recursion coefficients are computed by error propagation.
\begin{table}[t] 
 \centering
  \begin{tabular}{llll}
   \hline\\
   \multicolumn{1}{c}{$k$} & 
   \multicolumn{1}{c}{$u_k$} & 
   \multicolumn{1}{c}{$M/\mbar(1/\Lmax)$} & 
   \multicolumn{1}{c}{$M/\mbar(1/\Lmax)$} \\[0.5ex]
   & & 
   \multicolumn{1}{c}{2/3--loop} & 
   \multicolumn{1}{c}{1/2--loop} \\[1.0ex]
   \hline\\
 0 & 4.61      & 1.274     & 1.267 \\
 1 & 3.032(16) & 1.296(6)  & 1.292 \\
 2 & 2.341(21) & 1.295(10) & 1.292 \\
 3 & 1.918(20) & 1.294(13) & 1.292 \\
 4 & 1.628(16) & 1.294(14) & 1.292 \\
 5 & 1.414(14) & 1.295(15) & 1.293 \\
 6 & 1.251(12) & 1.297(16) & 1.295 \\
 7 & 1.121(10) & 1.298(17) & 1.297 \\
 8 & 1.017(10) & 1.299(17) & 1.298 \\[1.0ex]
   \hline
  \end{tabular} 
 \caption{Values for $M/\mbar(1/\Lmax)$ from \eq{Mombar}.}
 \label{t_Mombar}
\end{table}

Using the coefficients $w_k$ in \eq{wcoeff} we compute
\bea
 \frac{M}{\mbar(1/\Lmax)} & = & w_k^{-1}\frac{M}{\mbar(1/L_k)} \,, \label{Mombar}
\eea
where the factor $M/\mbar(1/L_k)$ is calculated from \eq{MRGI} with
$\gbar^2=u_k$ by employing the perturbative expressions for the $\tau$-- and
$\beta$--functions at 2-- respectively 3--loop order.
The results are shown in \tab{t_Mombar}. They have a remarkable
stability in the coupling\footnote{
The deviation in the case $k=0$ is due to the difference between the
perturbative and the non--perturbative values of $\sigma(u)$ at large $u$
(see \cite{DellaMorte:2004bc}).}
$u_k$ and we take $k=6$ as our result
\bea
 \frac{M}{\mbar(\mu)} & = & 1.297(16) \quad \mbox{at} \quad
 \mu \,=\, 1/\Lmax \,. \label{MombarLmax}
\eea
We emphasize that from \eq{MRGI} it is evident that $\sp(u)$ and
$M/\mbar(\mu)$ are flavor independent. Moreover, since the continuum limit
has been taken any regularization dependence has been removed from the result
\eq{MombarLmax}.
\begin{figure}[t]
 \begin{center}
     \includegraphics[width=13cm]{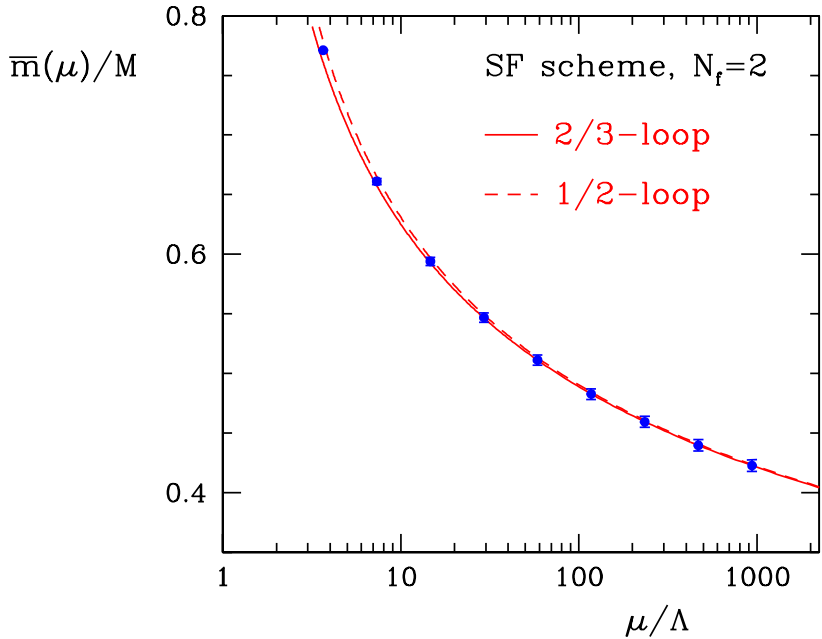}
 \end{center}
 \vspace{-2.5cm}
 \caption{The non--perturbative running of $\mbar$.
 2/3--loop refers to the 2--loop
 $\tau$-function and 3--loop $\beta$--function, analogously 1/2--loop.}
 \label{f_run}
\end{figure}

Finally, in \fig{f_run} we plot the non--perturbative running of the
renormalized mass. 
For $\mu/\Lambda \,=\, 1/(L_k\Lambda)$, $k=0,1,\ldots,8$, we plot the points
$\mbar(1/L_k)/M$ obtained from \eq{Mombar} using the result \eq{MombarLmax}.
The physical scale $\Lambda$ is here implicitly determined through
$\ln(\Lambda\Lmax)=-1.298(58)$ obtained from the recursion \eq{ucoeff}.
In the plot we neglect the overall uncertainties of $\mbar(1/\Lmax)/M$
and $\ln(\Lambda\Lmax)$,
since they would simply change the scales on the plot axes.
The errors of the points in \fig{f_run} come from the coefficients $w_k$.
Together with the non--perturbative points we show the perturbative curves
that are obtained from \eq{LAMBDA} and \eq{MRGI} by using
the perturbative expressions for the $\tau$-- and
$\beta$--functions at 1--, 2-- respectively 2--, 3--loop order.
The non--perturbative and perturbative running are very close down to the
smallest energies that were accessible in our simulations.
We remark that this statement explicitly refers to the special SF renormalization
scheme considered here.

\section{Estimate of the strange quark mass}
\label{mref}

\subsection{Complete $\zM$ for the improved Wilson discretization}
\label{Wilson}

We now derive the second factor in \eq{e:zMsplit} for a few values of the
lattice spacing or respectively the bare coupling. As emphasized before,
this contribution is non--universal and in the form given it will be valid only 
for our action of non--perturbatively improved Wilson fermions with plaquette
gauge action and $\csw$ as specified in \cite{Jansen:1998mx}. For $\ZA$ and $\cA$
we employ the values recently given and parameterized in \cite{DellaMorte:2005rd} 
and \cite{DellaMorte:2005se}. It remains to compute $\ZP(g_0,\Lmax/a)$ 
for the desired values of the bare
coupling $g_0$, here given by $\beta=5.2,5.29,5.4$.
The scale $\Lmax$ is fixed by $\gbar^2(\Lmax)=4.61$, where the universal factor 
of $\zM$ is now known, \eq{MombarLmax}.
Our basis are the simulation results summarized in \tab{t_ZP}.
\begin{table}[t] 
 \centering
  \begin{tabular}{lllll}
   \hline\\
   \multicolumn{1}{c}{$\beta$}  & 
   \multicolumn{1}{c}{$\kappa$} & 
   \multicolumn{1}{c}{$L/a$} & 
   \multicolumn{1}{c}{$\gbar^2$} & 
   \multicolumn{1}{c}{$\ZP$} \\[1.0ex]
   \hline\\
5.20 & 0.13600 & 4 & 3.65(3)   & 0.55188(51)  \\
5.20 & 0.13600 & 6 & 4.61(4)   & 0.47876(47)  \\[0.5ex]

5.29 & 0.13641 & 4 & 3.394(17) & 0.57643(50)  \\
5.29 & 0.13641 & 6 & 4.279(37) & 0.50739(58)  \\ 
5.29 & 0.13641 & 8 & 5.65(9)   & 0.45799(75)  \\[0.5ex]

5.40 & 0.13669 & 4 & 3.188(24) & 0.60269(65)  \\
5.40 & 0.13669 & 6 & 3.861(34) & 0.53681(54)  \\
5.40 & 0.13669 & 8 & 4.747(63) & 0.49116(66)  \\[1.0ex]
   \hline
  \end{tabular} 
 \caption{Results for $\ZP$, $c_t$ set to 2--loop value. 
 The values of $\gbar^2$ are from \cite{DellaMorte:2004bc}.
 The hopping parameters $\kappa$ are set to the critical ones $(\kappa_c)$ of
 \cite{Gockeler:2004rp}.}
 \label{t_ZP}
\end{table}

While the simulation at the largest bare coupling is exactly at the target value
for $\gbar^2$,
the two other series of simulations require a slight interpolation.
This has been done using a fit ansatz motivated by \eq{MRGI}
\bea
 \ln(\ZP) & = & c_1 + c_2 \ln(\gbar^2) \label{logfitZP}
\eea
to interpolate $\ZP$ between two values of $\gbar^2$ straddling $4.61$. 
The fit takes into account the (independent) errors of both $\ZP$ and $\gbar^2$.
The fit error
is then augmented by the difference between the fit result from \eq{logfitZP} and the
result from a simple two point linear interpolation in $\gbar^2$.
The values of the coefficient $c_2$ in the fit \eq{logfitZP} are found to be
$-0.369(25)$ at $\beta=5.29$ and $-0.430(34)$ at $\beta=5.40$,
which are not far from $-d_0/(2b_0)=-0.4138$.

The resulting numbers for $\ZP$ and $\zM$ are summarized in \tab{t_ZM}.
The first error for $\zM$ comes from the error of the factor $\ZA/\ZP$.
The second error is the 1.2\% uncertainty in the universal factor $M/\mbar$
and should be added in quadrature to the quark mass error after the continuum limit
has eventually been taken.

\begin{table}[t] 
 \centering
  \begin{tabular}{lll}
   \hline\\
   \multicolumn{1}{c}{$\beta$}  & 
   \multicolumn{1}{c}{$\ZP$} & 
   \multicolumn{1}{c}{$\zM$} \\[1.0ex]
   \hline\\
5.20 & 0.47876(47) & 1.935(33)(24) \\
5.29 & 0.4936(34)  & 1.979(25)(24) \\
5.40 & 0.4974(33)  & 2.001(29)(25) \\
   \hline
  \end{tabular} 
 \caption{Results for $\ZP$ and finally $\zM$ for three coupling values.
          The action to which it refers is detailed in the text.}
 \label{t_ZM}
\end{table}

\subsection{The reference quark mass}
\label{submref}

As announced in the introduction
we next compute the reference quark mass $\Mref$
producing a pseudoscalar with the mass of the Kaon in our simulated 
two-flavour theory.
The  hopping parameter $\kref$
is tuned to keep the pseudoscalar mass $\mps$ in the relation
\bea
 (r_0(\kappa_c)\mps(\kref))^2 & = & (r_0\mK)^2 \,=\, 1.5736 \,, \label{kref}
\eea
where $r_0$ is the scale extracted from the static quark potential
\cite{Sommer:1993ce}.
This value corresponds to the K--meson mass $\mK^2=(495\MeV)^2$
for $r_0=0.5\fm$ \cite{Garden:1999fg}.
Using the data for $r_0(\kappa)$, $\mps(\kappa)$ and $\kappa_c$
available in Ref. \cite{Gockeler:2004rp} 
and the extrapolations to the chiral limit $r_0(\kappa_c)$ of Ref.
\cite{DellaMorte:2004bc},
we can safely perform a slight extrapolation to the value $\kappa=\kref$ defined in 
\eq{kref}. We employ the following fit ansatz for $r_0\mps$ as a function of
the bare subtracted quark mass $a\mq$
\bea
 \frac{(r_0(\kappa_c)\mps(\kappa))^2}{a\mq} & = & e_1 + e_2\cdot a\mq \,,
\eea
with fit coefficients $e_1$ and $e_2$. The results of the fits are
$\kref=0.135680(30)$ at $\beta=5.20$,
$\kref=0.136018(27)$ at $\beta=5.29$ and
$\kref=0.136293(25)$ at $\beta=5.40$.

The PCAC masses $a\mref$ have been computed in simulations at $\kappa=\kref$
and the results are summarized in \tab{t_ms}. 
In these simulations the parameter $\theta$ in \eq{SFparams} has been set to zero.
Barring cutoff effects the PCAC mass
is independent of the time $x_0$ at which the right hand side of \eq{PCACmass} is
evaluated. Therefore in \tab{t_ms} we average over a range $t_1:t_2$ of $x_0$
values around $x_0=L/2$, where the PCAC mass has a plateau.
We keep $t_1-t_2$ roughly constant in physical units.
We see a significant gain in statistical precision due to the averaging compared
to taking only the time $x_0=L/2$.
The error analysis of derived observables like the (averaged) PCAC mass and $\ZP$
has been done with the method of Ref. \cite{Wolff:2003sm}.

The cutoff effects in $\mref$ also depend on the volume.
In fact, at a lattice spacing of $0.1\fm$ this dependence
is rather strong \cite{Sommer:2003ne}. Thus one has to define the
PCAC mass for a fixed volume in order to ensure that
cutoff effects disappear smoothly as O($a^2$). Furthermore
it is preferable to choose $L$ relatively large. We
choose $L\gtrsim1.5\fm$, where the volume dependence
can be neglected. Indeed, we compared
results from a simulation at $\beta=5.2$, $\kappa=0.1355$, $c_{\rm sw}=2.02$
on a $L/a=16$ lattice with those from the JLQCD collaboration\footnote{
We are grateful to Takashi Kaneko for providing us with data for this comparison.}
obtained at the same parameters on a $20^3\times48$ lattice \cite{Aoki:2002uc}.
We could check that
at the level of 1.6\% statistical precision in our simulation,
the PCAC masses agree. We conclude that our volume is
large enough at $\beta=5.2$ for the volume dependence to be negligible.
Therefore at the higher $\beta$ values we
choose approximately matched (or larger) physical volumes.
\begin{table}[t] 
 \centering
  \begin{tabular}{ccccccc}
   \hline\\
   \multicolumn{1}{c}{$\beta$}  & 
   \multicolumn{1}{c}{$\kref$} & 
   \multicolumn{1}{c}{$L/a$} & 
   \multicolumn{1}{c}{$t_1/a:t_2/a$} & 
   \multicolumn{1}{c}{$a\mref$} & 
   \multicolumn{1}{c}{$\Mref [\MeV]$} &
   \multicolumn{1}{c}{$\Mref [\MeV]$} \\[0.5ex] 
   & & & & & & 
   \multicolumn{1}{c}{using $\ZAc$} \\[1.0ex]
   \hline\\
5.20 & 0.135680 & 16 & 7:9   & 0.01410(30) & 58.7(3.2) & 69.5(3.7) \\[0.5ex]

5.29 & 0.136018 & 16 & 6:10  & 0.01352(28) & 63.5(3.2) & 69.1(3.6) \\[0.5ex]

5.40 & 0.136293 & 24 & 10:14 & 0.01300(18) & 72.0(2.7) & 76.2(2.7) \\[1.0ex]
   \hline
  \end{tabular} 
 \caption{Results for the PCAC masses and the RGI reference quark masses,
 the latter converted to $\MeV$ assuming $r_0=0.5\fm$. Two definitions of
 $\ZA$ are used \cite{DellaMorte:2005rd}.}
 \label{t_ms}
\end{table}

The second to last column of \tab{t_ms} shows our results
for the reference RGI quark mass
\be
\Mref = \zM \mref \label{Mref}.
\ee
We do see large cutoff effects.
At $\beta=5.2$ similarly large cutoff effects
have been observed in other quantities as well
\cite{Sommer:2003ne,DellaMorte:2004hs,DellaMorte:2005rd}. Moreover
the lattice spacing in our simulations changes by 30\% only when going
from $\beta=5.2$ to $\beta=5.4$. 
We therefore do not attempt (and discourage)
any elaborate continuum extrapolation
based on this range of lattice spacings. Instead we use a conservative estimate
of the continuum value of $\Mref$ by taking its value at our largest 
$\beta=5.4$ with the difference to the value at $\beta=5.2$ added as systematic
error
\bea
 \Mref & = & 72(3)(13)\MeV \,. \label{Mrefcont}
\eea
This result makes it clear that at present
the systematic error is dominating over the
statistical one.

The last column of \tab{t_ms} shows the results for an alternative estimate of
$\Mref$, which differs only by the use of the renormalization factor $\ZAc$
\cite{DellaMorte:2005rd}. We observe that the difference in $\Mref$ between the
two choices for $\ZA$ is compatible with an $a^2$--behavior.
Actually, the data obtained with $\ZAc$ have a weaker $a$--dependence
but this effect is essentially due to the change at the coarsest
lattice spacing ($\beta=5.2$). 

\subsection{The strange quark mass}
\label{subms}

To make contact with physics we now assume that \eq{Mrefcont}
also holds within errors in a theory with three degenerate flavors,
which can then be related to the strange quark in QCD by 
chiral perturbation theory.
This assumption is supported by the fact that
the value of $\Mref$ \eq{Mrefcont} is the same within errors as in the $\Nf=0$
theory \cite{Garden:1999fg}.

At lowest order in chiral perturbation theory, disregarding the
electromagnetic interaction, the formula
\cite{Gell-Mann:1968rz,Gasser:1982ap,Gasser:1984gg}
\bea
 \mK^2 \,=\,
 \frac{1}{2}(m_{{\rm K}^+}^2 + m_{{\rm K}^0}^2) & = & (\hat{M}+\Ms) \, \BRGI 
 \label{mKsq} \,,
\eea
where $\hat{M}=1/2(\Mu+\Md)$, holds.
Here $\Mu$, $\Md$, $\Ms$ are the up, down, strange RGI quark masses and
$\BRGI$ is a constant of the chiral Lagrangian.
\eq{mKsq} implies for degenerate quarks of RGI mass $\Mref$
(defined according to \eq{kref})
\bea
 \mK^2 & = & 2 \, \Mref \, \BRGI \,. \label{mrefchipt}
\eea
Therefore the relation $\Mref=(\hat{M}+\Ms)/2$ holds at lowest order
in chiral perturbation theory and
using $\Ms/\hat{M}=24.4(1.5)$ \cite{Leutwyler:1996qg} gives
\bea
 \Ms & \approx & 48/25\,\Mref \,. \label{mschipt}
\eea
Corrections to \eq{mschipt} are expected to be small in chiral perturbation theory
\cite{Gasser:1983yg,Bijnens:1994qh,Garden:1999fg,Amoros:2001cp}
and anyway below the accuracy that we will reach here for our result.
In addition in the quenched approximation it was found that the dependence
of pseudoscalar masses $\mps$, at fixed average quark mass,
on the difference of quark masses
is rather small \cite{Garden:1999fg}. We therefore assume
\eq{mschipt} to hold also with dynamical quarks.

At this point we are ready to give an estimate of the 
continuum value of the RGI strange quark mass
\bea
 \Ms & = & 138(5)(26)\MeV \,, \label{Ms}
\eea
by combining \eq{mschipt} with \eq{Mrefcont}.
Equivalently to the determination of $\Mref$, from \eq{mrefchipt} and \eq{Mrefcont}
we get
\bea
 \BRGI & = & 1.70(38)\GeV \,. \label{BRGI}
\eea
\begin{table}[t] 
 \centering
  \begin{tabular}{cccc}
   \hline\\
   $\mu [\GeV]$ & \multicolumn{3}{c}{$\mbar^{\MSb}(\mu)/M$} \\[0.5ex]
   & 2--loop & 3--loop & 4--loop \\[1.0ex]
   \hline\\
   1.0  & 0.7979 & 0.8364 & 0.8469 \\
   2.0  & 0.6824 & 0.6984 & 0.7013 \\
   4.0  & 0.6110 & 0.6197 & 0.6209 \\
   8.0  & 0.5608 & 0.5662 & 0.5668 \\
   90.0 & 0.4571 & 0.4588 & 0.4589 \\[1.0ex]
   \hline
  \end{tabular}
 \caption{Factors to convert the renormalization group invariant mass into the
 $\MSb$ scheme at scale $\mu$, for $\Lambda_{\MSb}^{(2)}=245(32)$ 
 \cite{DellaMorte:2004bc} and $\Nf=2$.}
 \label{t_moM}
\end{table}
\vspace{-0.5cm}

The renormalized strange quark mass in the $\MSb$ scheme at the renormalization
scale $\mu$ is obtained by multiplying $\Ms$ with the conversion factor
$\mbar^{\MSb}(\mu)/M$.
The latter is computed perturbatively
by numerical integration of \eq{LAMBDA} and \eq{MRGI} and is listed in \tab{t_moM}
for different choices of $\mu$.
The $n=2,3,4$--loop approximations of the $\beta$ and $\tau$ functions 
in the $\MSb$ scheme (for $\Nf=2$) are used.
The coefficients at 4--loop have been computed in
\cite{vanRitbergen:1997va,Chetyrkin:1997dh,Vermaseren:1997fq,Czakon:2004bu}.
To set $\mu$ in physical units the result
$\Lambda_{\MSb}^{(2)}=245(32)$ of Ref. \cite{DellaMorte:2004bc} is used.
The uncertainty in the $\Lambda$ parameter translates into a 2.8\% uncertainty in
$\mbar^{\MSb}(\mu)/M$ at $\mu=2\GeV$ and 4.9\% at $\mu=1\GeV$ (with 4--loop
evolution). Taking this into account our estimate for the
$\MSb$ strange quark mass is
\bea
 \mbars^{\MSb}(\mu) & = & 97(22)\MeV \quad \mbox{at} \quad \mu=2\GeV \,.
 \label{mbars}
\eea
\vspace{-1.0cm}
\section{Conclusions and outlook}
\label{conclusions}

We have presented a fully non--perturbative renormalization of the quark mass
in two flavor QCD, with renormalization conditions specified at zero quark mass
in the Schr{\"o}dinger Functional scheme.
Simulations were performed with O($a$) improved Wilson quarks.
Our main results are the running of the quark mass in \fig{f_run} and the
factor $M/\mbar(\mu)$ relating the quark mass at a specified low energy scale $\mu$
with the RGI quark mass in the continuum limit, \eq{MombarLmax}.

In order to obtain the renormalized strange quark mass in physical units
an appropriate hadronic scheme has to be defined. By using
existing data on the pseudoscalar masses and the scale $r_0$ from Ref.
\cite{Gockeler:2004rp} we were able to determine the RGI reference quark mass
$\Mref$ defined through \eq{Mref} and \eq{kref}, at three
lattice spacings in the approximate range $0.092\ldots0.071\fm$. Even in this small
range we see distinct cutoff effects. At the largest lattice spacing
large cutoff effects have been observed elsewhere
\cite{Sommer:2003ne,DellaMorte:2004hs,DellaMorte:2005rd}. These facts make it
impossible to perform a systematic
continuum extrapolation. Nevertheless we can give a conservative
estimate of the continuum value in \eq{Mrefcont}.
We emphasize that the error in \eq{Mrefcont} is dominated by the systematic
uncertainty of this step. 

Using a number of additional reasonable assumptions
we convert $\Mref$ to the strange quark mass in the $\MSb$ scheme \eq{mbars}.
A definite clarification that all of
these lead to negligible errors will require some
future work.
\begin{figure}[t]
 \begin{center}
     \includegraphics[width=12cm]{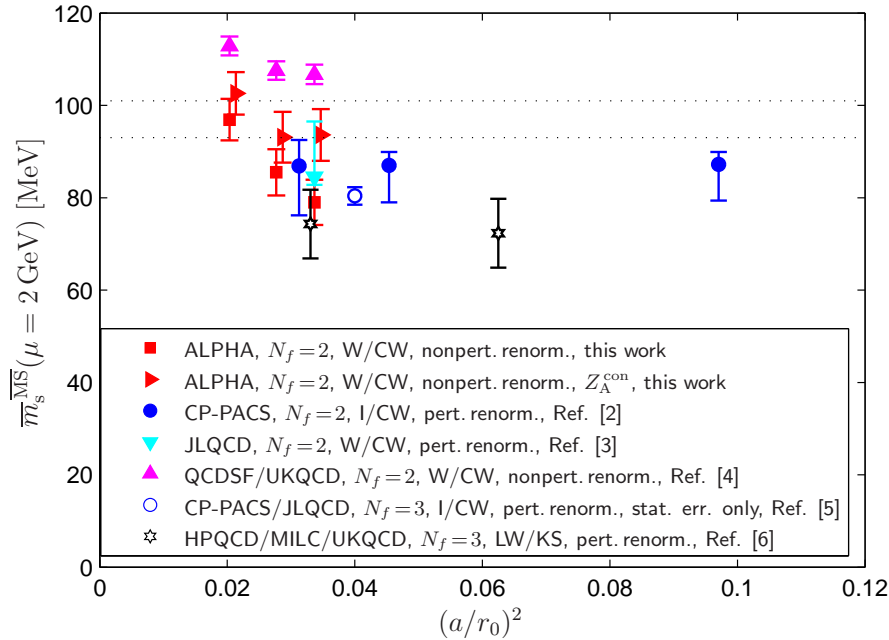}
 \end{center}
 \vspace{-0.5cm}
 \caption{Summary of the strange quark mass data from lattice simulations.
 In the legend the discretizations used are indicated in the form
 gauge action/fermion action. The dictionary reads:
 W: Wilson gauge action; I: Iwasaki gauge action; 
 LW: 1--loop tadpole improved L{\"u}scher--Weisz gauge action; 
 CW: Wilson--clover fermion action; KS: Asqtad staggered fermion action.
 The dotted lines represent the quenched result \cite{Garden:1999fg}.}
 \label{f_allms}
\end{figure}

In \fig{f_allms} we summarize the most recent results for the strange quark
mass from lattice simulations with two
\cite{AliKhan:2001tx,Aoki:2002uc,Gockeler:2004rp} and two plus one
\cite{Ishikawa:2004xq,Aubin:2004ck}
dynamical quarks. The physical K--meson mass is always used as input
although in some cases this is employed in a partially quenched setup where
valence and sea quark masses differ, while we compute $\Mref$ and then use
chiral perturbation theory to connect to the physical theory.
In the legend we emphasize whether nonperturbative or
perturbative renormalization has been used as well as the adopted discretizations.
It appears that perturbative renormalization leads to rather small values for the
strange quark mass.\footnote{
Note that the use of lowest order chiral perturbation theory in our work is not
likely to be a significant source of difference to the other computations, since
they do not report large deviations from lowest order chiral perturbation theory,
see for example Ref. \cite{Aoki:2002uc}.}
We plot the results of our present work using two
renormalization conditions for $\ZA$ that differ at O($a^2$). The red triangles
are obtained with $\ZAc$ \cite{DellaMorte:2005rd} and are slightly displaced
for clarity. This comparison gives a flavor of the quite large cutoff effects at
$\beta=5.2$, our coarsest lattice spacing.
The dotted lines in \fig{f_allms} mark the quenched result of
Ref. \cite{Garden:1999fg}.
Given the present status, illustrated in \fig{f_allms}, it appears
hard to claim a definite dependence of $\ms$ on the number
of dynamical fermions, even between $\nf=0$ and $\nf=2$.

Our present result for the renormalized strange quark mass should be
improved by simulating at a finer lattice spacing. We are confident that this
will be possible in the near future. There are promising improvements in algorithms
\cite{Luscher:2003vf,Luscher:2004rx,Urbach:2005ji}, and a new machine apeNEXT
is becoming available to us \cite{Bodin:2001hn,Bodin:2005gg}.
Moreover the determination
of the lattice spacing can be improved by, for example, computing the
renormalized strange quark mass in units of the K--meson decay constant $\FK$.
There are first indications
that the SF is an efficient setup for this low--energy computation
(see also \cite{Guagnelli:1999zf}).

\bigskip

{\bf Acknowledgement.}
We are grateful to Gerrit Schierholz for communicating results of Ref.
\cite{Gockeler:2004rp}.
We further thank NIC/DESY for allocating computing resources on the APEmille machine
to this project and the APE collaboration
and the staff of the computer center at DESY, Zeuthen for their support.

The computation of the renormalized quark mass is part of project B2 of the
SFB Transregio 9 ``Computational Particle Physics'' and has also been
supported by the Deutsche Forschungsgemeinschaft (DFG)
in the Graduiertenkolleg GK 271 as well as by the 
European Community's Human Potential Programme under contract HPRN-CT-2000-00145.

\begin{appendix}
\section{Simulation results for $\ZP$}
\label{app_zp}

\begin{table}[ht!] 
 \centering
  \begin{tabular}{rrlllll}
   \hline\\
   \multicolumn{1}{c}{$L/a$} & 
   \multicolumn{1}{c}{$\beta=6/g_0^2$} & 
   \multicolumn{1}{c}{$\kappa$} & 
   \multicolumn{1}{c}{$u=\gbar^2(L)$} & 
   \multicolumn{1}{c}{$\ZP(g_0,L/a)$} & 
   \multicolumn{1}{c}{$\ZP(g_0,2L/a)$} & 
   \multicolumn{1}{c}{$\SP(u,a/L)$} \\[1.0ex]
   \hline\\[0.5ex]
 6  & 9.5000   & 0.1315322 & 0.9793(11) & 0.8265(4) & 0.7940(7) & 0.9606(10) \\
 8  & 9.7341   & 0.1313050 & 0.9807(17) & 0.8195(5) & 0.7903(8) & 0.9643(11) \\
12  & 10.05755 & 0.1310691 & 0.9792(33) & 0.8095(5) & 0.7828(10) & 0.9670(14) \\[2.0ex]

 6  & 8.5000   & 0.1325094 & 1.1814(15) & 0.7985(5) & 0.7606(16) & 0.9526(21) \\
 8  & 8.7223   & 0.1322907 & 1.1818(29) & 0.7896(9) & 0.7527(9) & 0.9533(14) \\
12  & 8.99366  & 0.1319754 & 1.1814(78) & 0.7793(5) & 0.7417(13) & 0.9517(19) \\[2.0ex]
                                
 6  & 7.5000   & 0.1338150 & 1.5031(25) & 0.7565(4) & 0.7115(13) & 0.9405(18) \\
 6  & 7.5420   & 0.1337050 & 1.5078(44) & 0.7604(4) & 0.7125(9) & 0.9370(13) \\
 8  & 7.7206   & 0.1334970 & 1.5077(43) & 0.7486(8) & 0.7045(17) & 0.9411(25) \\
12  & 8.02599  & 0.1330633 & 1.503(12)  & 0.7401(11) & 0.6967(12) & 0.9414(22) \\[2.0ex]

 6  & 6.6085  & 0.1352600  & 2.0146(56) & 0.7025(5) & 0.6406(7) & 0.9118(12) \\
 8  & 6.8217  & 0.1348910  & 2.014(10)  & 0.6921(11) & 0.6366(10) & 0.9198(21) \\
12  & 7.09300 & 0.1344320  & 2.014(20)  & 0.6827(10) & 0.6236(15) & 0.9134(27) \\[2.0ex]

 6  & 6.1330  & 0.1361100  & 2.488(11)  & 0.6584(7) & 0.5824(18) & 0.8845(29) \\
 8  & 6.3229  & 0.1357673  & 2.479(13)  & 0.6500(7) & 0.5755(21) & 0.8854(33) \\
12  & 6.63164 & 0.1352270  & 2.479(25)  & 0.6431(10) & 0.5716(18) & 0.8888(33) \\[2.0ex]
                                
 6  & 5.6215  & 0.1366650  & 3.326(20)  & 0.5857(12) & 0.4898(25) & 0.8363(47) \\
 8  & 5.8097  & 0.1366077  & 3.334(19)  & 0.5810(12) & 0.4864(24) & 0.8371(45) \\
12  & 6.11816 & 0.1361387  & 3.334(49)  & 0.5817(14) & 0.4888(28) & 0.8403(55) \\[1.0ex]
   \hline
  \end{tabular} 
 \caption{Results for the step scaling function $\SP$.}
 \label{t_SigmaP}
\end{table}
\noindent
In \tab{t_SigmaP} we collect the bare parameters and results
of our simulations to compute $\ZP$.
Simulations on $L/a$ and $2L/a$ lattices are required to extract the
step scaling function $\SP$ \eq{ssfZP}.
At the three lowest couplings $\gbar^2(L)$ simulations
have been performed using the 1--loop value of $\ct$
\cite{Luscher:1992an}, except for
$L/a=6$, $\beta=7.5420$ and $L/a=8$, $\beta=7.7206$. For the latter parameters
and the larger couplings the 2--loop value of $\ct$ \cite{Bode:1999sm}
has been used.
At the third lowest coupling $u\approx1.5$ we checked at $L/a=6$ that there
is no significant difference in $\SP$ using the 1-- or 2--loop value for $\ct$,
as it is shown in \fig{f_SigmaP}.

To the statistical error of $\SP$ we added in quadrature the error 
due to the uncertainty in the coupling $u$. The latter was estimated using
the 1--loop result $-\ln(2)d_0$ for the derivative of $\SP$ with respect to $u$.
This correction is tiny, it increases the errors of $\SP$ by at most $5\%$ at
the largest coupling and lattice.

\section{About the algorithm}
\label{algorithm}
\begin{figure}[t]
 \begin{center}
     \includegraphics[width=12cm]{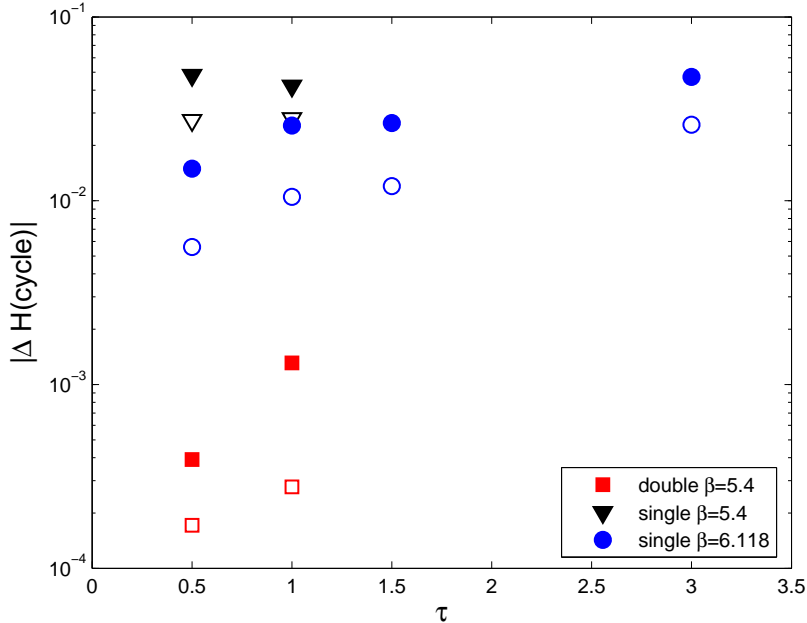}
 \end{center}
 \vspace{-1.0cm}
 \caption{Reversibility violations in the HMC as a function of the
 trajectory length $\tau$ used in the reversibility tests. Empty symbols
 represent the average and filled symbols the maximal value of the
 Hamiltonian violation $\DHc$ \cite{DellaMorte:2003jj}.}
 \label{f_rev}
\end{figure}

Our simulations have been performed using the Hybrid Monte Carlo algorithm with
two pseudo--fermion fields as proposed by M. Hasenbusch
\cite{Hasenbusch:2001ne,Hasenbusch:2002ai,DellaMorte:2003jj}.
Here we extend the study on algorithmic precision presented in Ref.
\cite{DellaMorte:2003jj}.

For our largest lattice volume $L/a=24$ at $\beta=5.4$ we investigated the issue
of reversibility.
We looked at the quantity $\DHc$ which measures the
difference of the Hamiltonian for a cyclic trajectory \cite{DellaMorte:2003jj}.
In \fig{f_rev} we present data for the average $\DHc$ (empty symbols) and
the maximum of $\DHc$ (filled symbols) on three sets of configurations.
Sticking to single precision arithmetic we found for a trajectory length $\tau=0.5$
(that we use in the production run) an average value for $\DHc$ 
(black triangles) that is larger by almost a factor five with
respect to the simulations at $\beta=6.118$
reported in Ref. \cite{DellaMorte:2003jj} (blue circles).
We therefore repeated the simulation at $\beta=5.4$ in double precision arithmetic.
At the same time we reduced the parameter $\epsilon^2$, defined\footnote{
We adopt here the commonly used convention to define $\epsilon^2$ as the square of
the ratio between the norm of the residue vector and the norm of the source vector.
It differs from the definition used in \cite{DellaMorte:2003jj}, where $\epsilon^2$
was normalized by the norm of the solution vector.}
as the requested accuracy in the conjugate gradient iteration,
from $\epsilon^2=10^{-11}$ (as used with single precision) to
$\epsilon^2=10^{-13}$. The results for double precision are
the red squares in \fig{f_rev}. By comparing single with double precision,
both the average and the maximal value of $\DHc$ decrease by two orders of
magnitude for double precision and these values are well below the values
observed in Ref. \cite{DellaMorte:2003jj}. Our statistics for the
reversibility checks consists of 18 configurations analyzed at $\beta=5.4$ (both
for single and double precision) and of 30 configurations analyzed at
$\beta=6.118$. On a subset of the $\beta=5.4$ configurations we checked that
within single precision by reducing $\epsilon^2$ to $10^{-13}$ (which one
would expect to be the lower bound in single precision) the roundoff error in the
Hamiltonian computation and consequently the value of $\DHc$ remain unchanged.

Finally, we emphasize that we found {\em no}
significant difference in the observables we computed at $\beta=5.4$
between the single and double precision simulations.
The number quoted in \tab{t_ms} for $\mref$ is from the simulation in double
precision.

\end{appendix}

\bibliography{mass}           %or whatever your .bib file is

\begin{thebibliography}{10}

\bibitem{Weinberg:1951ss}
S. Weinberg,
\newblock Phys. Rev. D8 (1973) 3497.

\bibitem{AliKhan:2001tx}
CP-PACS, A. Ali~Khan et~al.,
\newblock Phys. Rev. D65 (2002) 054505, hep-lat/0105015.

\bibitem{Aoki:2002uc}
JLQCD, S. Aoki et~al.,
\newblock Phys. Rev. D68 (2003) 054502, hep-lat/0212039.

\bibitem{Gockeler:2004rp}
QCDSF, M. G{\"o}ckeler et~al.,
\newblock (2004), hep-ph/0409312.

\bibitem{Ishikawa:2004xq}
CP-PACS, T. Ishikawa et~al.,
\newblock Nucl. Phys. Proc. Suppl. 140 (2005) 225, hep-lat/0409124.

\bibitem{Aubin:2004ck}
HPQCD, C. Aubin et~al.,
\newblock Phys. Rev. D70 (2004) 031504, hep-lat/0405022.

\bibitem{Capitani:1998mq}
ALPHA, S. Capitani, M. L{\"u}scher, R. Sommer and H. Wittig,
\newblock Nucl. Phys. B544 (1999) 669, hep-lat/9810063.

\bibitem{DellaMorte:2004bc}
ALPHA, M. Della~Morte et~al.,
\newblock Nucl. Phys. B713 (2005) 378, hep-lat/0411025.

\bibitem{Knechtli:2002vp}
ALPHA, F. Knechtli et~al.,
\newblock Nucl. Phys. Proc. Suppl. 119 (2003) 320, hep-lat/0209025.

\bibitem{Knechtli:2003mq}
ALPHA, F. Knechtli et~al.,
\newblock Nucl. Phys. Proc. Suppl. 129 (2004) 814, hep-lat/0309074.

\bibitem{DellaMorte:2003jj}
ALPHA, M. Della~Morte et~al.,
\newblock Comput. Phys. Commun. 156 (2003) 62, hep-lat/0307008.

\bibitem{DellaMorte:2005se}
M. Della~Morte, R. Hoffmann and R. Sommer,
\newblock JHEP 03 (2005) 029, hep-lat/0503003.

\bibitem{DellaMorte:2005rd}
M. Della~Morte, R. Hoffmann, F. Knechtli, R. Sommer and U. Wolff,
\newblock JHEP 07 (2005) 007, hep-lat/0505026.

\bibitem{Rolf:2002gu}
ALPHA, J. Rolf and S. Sint,
\newblock JHEP 12 (2002) 007, hep-ph/0209255.

\bibitem{Dougall:2004hx}
UKQCD, A. Dougall, C.M. Maynard and C. McNeile,
\newblock Nucl. Phys. Proc. Suppl. 140 (2005) 428, hep-lat/0409089.

\bibitem{Heitger:2003nj}
ALPHA, J. Heitger and R. Sommer,
\newblock JHEP 02 (2004) 022, hep-lat/0310035.

\bibitem{Garden:1999fg}
ALPHA, J. Garden, J. Heitger, R. Sommer and H. Wittig,
\newblock Nucl. Phys. B571 (2000) 237, hep-lat/9906013.

\bibitem{Gasser:1984gg}
J. Gasser and H. Leutwyler,
\newblock Nucl. Phys. B250 (1985) 465.

\bibitem{Leutwyler:1996qg}
H. Leutwyler,
\newblock Phys. Lett. B378 (1996) 313, hep-ph/9602366.

\bibitem{Amoros:2001cp}
G. Amoros, J. Bijnens and P. Talavera,
\newblock Nucl. Phys. B602 (2001) 87, hep-ph/0101127.

\bibitem{Wittig:2002ux}
H. Wittig,
\newblock Nucl. Phys. Proc. Suppl. 119 (2003) 59, hep-lat/0210025.

\bibitem{Kaplan:1986ru}
D.B. Kaplan and A.V. Manohar,
\newblock Phys. Rev. Lett. 56 (1986) 2004.

\bibitem{Gell-Mann:1968rz}
M. Gell-Mann, R.J. Oakes and B. Renner,
\newblock Phys. Rev. 175 (1968) 2195.

\bibitem{Sint:1998iq}
ALPHA, S. Sint and P. Weisz,
\newblock Nucl. Phys. B545 (1999) 529, hep-lat/9808013.

\bibitem{Jansen:1995ck}
K. Jansen et~al.,
\newblock Phys. Lett. B372 (1996) 275, hep-lat/9512009.

\bibitem{Luscher:1998pe}
M. L{\"u}scher,
\newblock (1998), hep-lat/9802029.

\bibitem{Symanzik:1981wd}
K. Symanzik,
\newblock Nucl. Phys. B190 (1981) 1.

\bibitem{Luscher:1985iu}
M. L{\"u}scher,
\newblock Nucl. Phys. B254 (1985) 52.

\bibitem{Luscher:1992an}
M. L{\"u}scher, R. Narayanan, P. Weisz and U. Wolff,
\newblock Nucl. Phys. B384 (1992) 168, hep-lat/9207009.

\bibitem{Sint:1993un}
S. Sint,
\newblock Nucl. Phys. B421 (1994) 135, hep-lat/9312079.

\bibitem{Sint:1995rb}
S. Sint,
\newblock Nucl. Phys. B451 (1995) 416, hep-lat/9504005.

\bibitem{Sint:1995ch}
S. Sint and R. Sommer,
\newblock Nucl. Phys. B465 (1996) 71, hep-lat/9508012.

\bibitem{Luscher:1996sc}
M. L{\"u}scher, S. Sint, R. Sommer and P. Weisz,
\newblock Nucl. Phys. B478 (1996) 365, hep-lat/9605038.

\bibitem{Jansen:1998mx}
ALPHA, K. Jansen and R. Sommer,
\newblock Nucl. Phys. B530 (1998) 185, hep-lat/9803017.

\bibitem{Sint:1997jx}
S. Sint and P. Weisz,
\newblock Nucl. Phys. B502 (1997) 251, hep-lat/9704001.

\bibitem{Guagnelli:2000jw}
ALPHA, M. Guagnelli et~al.,
\newblock Nucl. Phys. B595 (2001) 44, hep-lat/0009021.

\bibitem{Bode:1999sm}
ALPHA, A. Bode, P. Weisz and U. Wolff,
\newblock Nucl. Phys. B576 (2000) 517, hep-lat/9911018.

\bibitem{Sommer:1993ce}
R. Sommer,
\newblock Nucl. Phys. B411 (1994) 839, hep-lat/9310022.

\bibitem{Wolff:2003sm}
ALPHA, U. Wolff,
\newblock Comput. Phys. Commun. 156 (2004) 143, hep-lat/0306017.

\bibitem{Sommer:2003ne}
ALPHA, R. Sommer et~al.,
\newblock Nucl. Phys. Proc. Suppl. 129 (2004) 405, hep-lat/0309171.

\bibitem{DellaMorte:2004hs}
ALPHA, M. Della~Morte, R. Hoffmann, F. Knechtli and U. Wolff,
\newblock Comput. Phys. Commun. 165 (2005) 49, hep-lat/0405017.

\bibitem{Gasser:1982ap}
J. Gasser and H. Leutwyler,
\newblock Phys. Rept. 87 (1982) 77.

\bibitem{Gasser:1983yg}
J. Gasser and H. Leutwyler,
\newblock Ann. Phys. 158 (1984) 142.

\bibitem{Bijnens:1994qh}
J. Bijnens, G. Ecker and J. Gasser,
\newblock (1994), hep-ph/9411232.

\bibitem{vanRitbergen:1997va}
T. van Ritbergen, J.A.M. Vermaseren and S.A. Larin,
\newblock Phys. Lett. B400 (1997) 379, hep-ph/9701390.

\bibitem{Chetyrkin:1997dh}
K.G. Chetyrkin,
\newblock Phys. Lett. B404 (1997) 161, hep-ph/9703278.

\bibitem{Vermaseren:1997fq}
J.A.M. Vermaseren, S.A. Larin and T. van Ritbergen,
\newblock Phys. Lett. B405 (1997) 327, hep-ph/9703284.

\bibitem{Czakon:2004bu}
M. Czakon,
\newblock Nucl. Phys. B710 (2005) 485, hep-ph/0411261.

\bibitem{Luscher:2003vf}
M. L{\"u}scher,
\newblock JHEP 05 (2003) 052, hep-lat/0304007.

\bibitem{Luscher:2004rx}
M. L{\"u}scher,
\newblock Comput. Phys. Commun. 165 (2005) 199, hep-lat/0409106.

\bibitem{Urbach:2005ji}
C. Urbach, K. Jansen, A. Shindler and U. Wenger,
\newblock (2005), hep-lat/0506011.

\bibitem{Bodin:2001hn}
F. Bodin et~al.,
\newblock Nucl. Phys. Proc. Suppl. 106 (2002) 173, hep-lat/0110197.

\bibitem{Bodin:2005gg}
ApeNEXT, F. Bodin et~al.,
\newblock Nucl. Phys. Proc. Suppl. 140 (2005) 176.

\bibitem{Guagnelli:1999zf}
ALPHA, M. Guagnelli, J. Heitger, R. Sommer and H. Wittig,
\newblock Nucl. Phys. B560 (1999) 465, hep-lat/9903040.

\bibitem{Hasenbusch:2001ne}
M. Hasenbusch,
\newblock Phys. Lett. B519 (2001) 177, hep-lat/0107019.

\bibitem{Hasenbusch:2002ai}
M. Hasenbusch and K. Jansen,
\newblock Nucl. Phys. B659 (2003) 299, hep-lat/0211042.

\end{thebibliography}
\bibliographystyle{h-elsevier}   %if you use h-elsevier.bst

\end{document}